\let\ifarxiv=\iftrue     % ARXIV VERSION
\pdfoutput=1

\ifarxiv

\documentclass[12pt,a4paper]{article}
\usepackage[a4paper,text={450pt,650pt},centering]{geometry}

\fi

\ifarxiv\else

\documentclass[11pt,a4paper]{article}
\usepackage{mathptmx}
\usepackage[a4paper,text={130mm,198mm}]{geometry}

\fi

\usepackage{amsmath,amssymb}
\usepackage[bookmarks=true,hyperfigures=true]{hyperref}
\usepackage{graphicx}
\usepackage[nosort]{cite}
\usepackage[T1]{fontenc}

\let\oldbfseries=\bfseries
\let\oldmdseries=\mdseries
\let\oldnormalfont=\normalfont
\renewcommand{\bfseries}{\oldbfseries\boldmath}
\renewcommand{\mdseries}{\oldmdseries\unboldmath}
\renewcommand{\normalfont}{\oldnormalfont\unboldmath}

\allowdisplaybreaks[3]

\numberwithin{equation}{section}

\usepackage[font=small,labelfont=bf,width=0.85\textwidth]{caption}

\providecommand{\hypersetup}[1]{}

\hypersetup{plainpages=false}
\hypersetup{pdfpagemode=UseNone}
\hypersetup{bookmarksnumbered=true}
\hypersetup{pdfstartview=FitH}
\hypersetup{colorlinks=false}
\hypersetup{citebordercolor={.5 1 .5}}
\hypersetup{urlbordercolor={.5 1 1}}
\hypersetup{linkbordercolor={1 .7 .7}}
\providecommand{\href}[2]{#2}
\providecommand{\arxivlink}[1]{\href{http://arxiv.org/abs/#1}{arxiv:#1}}

\newcommand{\alg}[1]{\mathfrak{#1}}
\newcommand{\Tr}{{\rm Tr \,}}

\begin{document}

%%%%%%%%%%%%%%%%%%%%%%%%%%%%%%%%%%%%%%%%%%%%%%%%%%%%%%%%%%%%%%%%%%%%%%%%%%%%%%%%
%%%%%%%%%%%%%%%%%%%%%%%%%%%%%%%%%%%%%%%%%%%%%%%%%%%%%%%%%%%%%%%%%%%%%%%%%%%%%%%%
% TITLE PAGE

\thispagestyle{empty}
\phantomsection
\addcontentsline{toc}{section}{Title}

\begin{flushright}\footnotesize%
\texttt{\arxivlink{1012.3990}},
\texttt{HU-Mathematik 2010-20, HU-EP-10/84, AEI-2010-176}
\vspace{1em}%
\end{flushright}

\begingroup\parindent0pt
\begingroup\bfseries\ifarxiv\Large\else\LARGE\fi
\hypersetup{pdftitle={Review of AdS/CFT Integrability, Chapter III.1: Bethe Ans\"atze and the R-Matrix Formalism}}%
Review of AdS/CFT Integrability, Chapter III.1:\\
Bethe Ans\"atze and the R-Matrix Formalism
\par\endgroup
\vspace{1.5em}
\begingroup\ifarxiv\scshape\else\large\fi%
\hypersetup{pdfauthor={Author(s)}}%
Matthias Staudacher
\par\endgroup
\vspace{1em}
\begingroup\itshape
Institut f\"ur Mathematik und Institut f\"ur Physik, Humboldt-Universit\"at zu Berlin\\
Johann von Neumann-Haus, Rudower Chaussee 25, 12489 Berlin, Germany\\
and\\
Max-Planck-Institut f\"ur Gravitationsphysik, Albert-Einstein-Institut\\
    Am M\"uhlenberg 1, 14476 Potsdam, Germany
\par\endgroup
\vspace{1em}
\begingroup\ttfamily
matthias@aei.mpg.de
\par\endgroup
\vspace{1.0em}
\endgroup

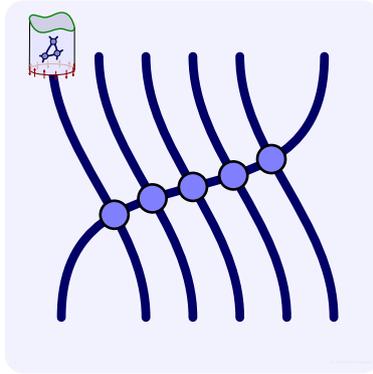
\begin{center}
\includegraphics[width=5cm]{TitleIII1.mps}%figure for your chapter
\vspace{1.0em}
\end{center}

\paragraph{Abstract:}
The one-dimensional Heisenberg XXX spin chain appears in a special limit of the AdS/CFT integrable system. We review various ways of proving its integrability, and discuss the associated methods of solution. In particular, we outline the coordinate and the algebraic Bethe ansatz, giving reference to literature suitable for learning these techniques. Finally, we speculate which of the methods might lift to the  exact solution of the AdS/CFT system, and sketch a promising method for constructing the Baxter Q-operator of the XXX chain. It allows to find the spectrum of the model using certain algebraic techniques, while entirely avoiding Bethe's ansatz.

\ifarxiv\else
\paragraph{Mathematics Subject Classification (2010):} 
82B23, 16T25
% http://www.ams.org/msc
\fi
\hypersetup{pdfsubject={MSC (2010): 82B23, 16T25}}%

\ifarxiv\else
\paragraph{Keywords:} 
Bethe ansatz, R-matrix, Q-operator
\fi
\hypersetup{pdfkeywords={...}}%

\newpage

%%%%%%%%%%%%%%%%%%%%%%%%%%%%%%%%%%%%%%%%%%%%%%%%%%%%%%%%%%%%%%%%%%%%%%%%%%%%%%%%
%%%%%%%%%%%%%%%%%%%%%%%%%%%%%%%%%%%%%%%%%%%%%%%%%%%%%%%%%%%%%%%%%%%%%%%%%%%%%%%%
% BODY

%%%%%%%%%%%%%%%%%%%%%%%%%%%%%%%%%%%%%%%%%%%%%%%%%%%%%%%%%%%%%%%%%%%%%%%%%%%%%%%%
\section{Introduction}
\label{sec:Intro}

Quantum integrability was discovered in 1931 by young postdoc Hans Bethe during a research stay in Rome. Interestingly, this happened while the general formalism of non-relativistic quantum mechanics was still being developed. Bethe took a look at a one-dimensional model for a metal, the so-called Heisenberg spin chain, whose Hamiltonian reads
\begin{equation}\label{hamiltonian}
{\rm \bf{H}}=4\, \sum_{l=1}^{L}\left( \frac{1}{4}-\vec S_{l} \cdot \vec S_{l+1}\right)
\qquad {\rm with} \qquad 
\vec S_l =\frac{1}{2}\, \vec \sigma_l\, ,
\end{equation}
where $\vec \sigma_l$ are the three Pauli matrices, i.e.~each $\vec S_l$ is a separate spin-$\tfrac{1}{2}$ representation of $\alg{su}(2)$. He managed to find, in a sense to be explained below, the {\it exact} solution of this model by making a clever educated guess for the ``wave function'' $|\psi\rangle$ of the spin chain system, that is for the eigenstates of the spectral problem
\begin{equation}\label{schroedinger}
{\rm \bf{H}} \cdot |\psi\rangle=E\,|\psi\rangle\, ,
\end{equation}
where $E$ are the energy eigenvalues. His original paper is still very readable today, and easily accessible either in its original German version \cite{Bethe:1931aa}, or its English translation, which is easily available on the internet. The last sentence of this masterpiece, just before the acknowledgements to Enrico Fermi and the sponsor of his visit, the Rockefeller foundation, indicates that Bethe was not quite aware how lucky he had been discovering quantum integrability in one dimension: He states with confidence that he was intending to generalize his method to the solution of higher dimensional lattices in a follow-up paper. We now know that this was bound to fail.

Bethe's discovery of the exact solution of the spectral problem \eqref{schroedinger} is of timeless beauty and extraordinary importance for condensed matter theory and mathematical physics. The method continued and continues to be relevant to a multitude of widely differing problems. The maybe latest reincarnation occured in the context of integrability in AdS/CFT, which is the focus of this review. As explained in the articles of part I of this volume, the Hamiltonian \eqref{hamiltonian} appears in ${\cal N}=4$ Yang-Mills theory in the scalar field subsector, where $\vec S \in \alg{su}(2) \subset \alg{su}(4) \subset \alg{psu}(2,2|4)$, as the one-loop approximation of the conformal dilatation generator ${\rm \bf{D}} \in \alg{su}(2,2) \subset \alg{psu}(2,2|4)$  
\begin{equation}\label{dilatation}
{\rm \bf{D}} =L+g^2\, {\rm \bf{H}}+{\cal O}(g^4)\, .
\end{equation}
To understand the meaning of these Lie algebras in the AdS/CFT context, please refer to the article \cite{chapSuperconf}. Here $g^2$ is related to the `t Hooft coupling constant $\lambda$ by $g^2=\frac{\lambda}{16 \pi^2}$. Note that we did not yet fully define the Hamiltonian in \eqref{hamiltonian}, since we so far did not state how to interpret $\vec S_{L+1}$. In other words, we need to specify the {\it boundary conditions} of the chain. For ${\cal N}=4$ we need periodic boundary conditions 
\begin{equation}\label{periodic}
\vec S_{L+1}:=\vec S_1
\, ,
\end{equation}
which is also the case originally solved by Bethe. It is relatively easy to see that ${\rm \bf{H}}$ in \eqref{hamiltonian} with \eqref{periodic} is rotationally invariant, i.e.~commutes with the total spin operator $\vec S$:
\begin{equation}\label{totalspin}
[ {\rm \bf{H}}, \vec S] =0
\quad {\rm where} \quad
\vec S=\sum_{l=1}^L \vec S_l
\, .
\end{equation}

\subsection{Understanding the problem}

Let us first understand the problem, before contemplating its solution. The first thing to grasp is that the ${\rm \bf{H}}$ in \eqref{hamiltonian} is just a simple matrix of size $2^L \times 2^L$. Why? Because the spin chain is composed of $L$ Pauli spins, each of which can be pointing either up $\uparrow$ or down $\downarrow$. Or, as in Bethe's paper,\footnote{%
An interesting historical question is why Bethe's left-right convention lost out to up-down.
}%
left $\leftarrow$ or right $\rightarrow$. In the AdS/CFT context, the two spin orientations correspond to two of the three possible complex scalar matter fields, say Z and X, and the spin chain is a local composite single trace operator. A basis for the configuration space of the chain is of size $2^L$. The eigenstates $|\psi\rangle$ must then be the appropriate linear combinations of these $2^L$ basis vectors. The proper mathematical concept to describe this set-up is the tensor product. Let us denote the state space of a single spin by $\mathbb{C}^2$. Then a basis of this two-dimensional complex (as quantum mechanics demands) vector space is
\begin{equation}\label{singlespin}
| \uparrow\, \rangle=Z=\binom{1}{0},
\qquad
| \downarrow\, \rangle=X=\binom{0}{1}.
\end{equation}
The full state space of the quantum spin chain is then 
\begin{equation}\label{quantumspace}
%{\mathcal V}=
\underbrace{
{\mathbb C}^2\otimes {\mathbb C}^2\otimes \cdots \otimes  
{\mathbb C}^2}_{L-\mbox{\scriptsize{times}}}
\ ,
\end{equation}
and $\vec S_l$ in \eqref{hamiltonian} means that this operator acts like the $2 \times 2$ identity matrix $\mathbb{I}_2$ on each copy of $\mathbb{C}^2$, except for the $l$-th one, where it acts as $\frac{1}{2}\, \vec \sigma$.
Tensor products can be confusing, and I recommend the very pedagogical introduction \cite{Nepomechie:1998jf} for a transparent and detailed explanation in the same context. To illustrate, let us study the simplest non-trivial example of this, the $L=2$ spin chain. In the basis 
\begin{equation}\label{L2basis}
\Big\{ | \uparrow\, \rangle \otimes | \uparrow\, \rangle,
| \uparrow\, \rangle \otimes | \downarrow\, \rangle,
| \downarrow\, \rangle \otimes | \uparrow\, \rangle,
| \downarrow\, \rangle \otimes | \downarrow\, \rangle \Big\}
,
\end{equation}
or short
\begin{equation}\label{L2basisalt}
\Big\{ | \uparrow\, \uparrow\, \rangle,
| \uparrow\, \downarrow\, \rangle,
| \downarrow\, \uparrow\, \rangle,
| \downarrow\, \downarrow\, \rangle \Big\}
,
\end{equation}
we have from \eqref{hamiltonian} (the reader may have to play a bit with Pauli's matrices to see this)
\begin{equation}\label{hamiltonianL2}
{\rm \bf{H}}=
\begin{pmatrix}
0 & 0 & 0 & 0 \\
0 & +4 & -4 & 0 \\
0 & -4 & +4 & 0 \\
0 & 0 & 0 & 0 \\
\end{pmatrix}
.
\end{equation}
In this simplest case, it is of course trivial to find the eigensystem of the matrix ${\rm \bf{H}}$. The (not fully normalized) eigenvectors are
\begin{equation}\label{L2beigenbasis}
\Big\{ | \uparrow\, \uparrow\, \rangle,
| \uparrow\, \downarrow\, \rangle + | \downarrow\, \uparrow\, \rangle,
| \uparrow\, \downarrow\, \rangle - | \downarrow\, \uparrow\, \rangle,
| \downarrow\, \downarrow\, \rangle \Big\}
,
\end{equation}
and in this basis the diagonalized Hamiltonian reads
\begin{equation}\label{L2eigenvalues}
%{\rm \bf{H}}=
\begin{pmatrix}
0 & 0 & 0 & 0 \\
0 & 0 & 0 & 0 \\
0 & 0 & 8 & 0 \\
0 & 0 & 0 & 0 \\
\end{pmatrix}
.
\end{equation}
It is useful to reorder the basis to
\begin{equation}\label{L2beigenbasisalt}
\Big\{ | \uparrow\, \uparrow\, \rangle,
| \uparrow\, \downarrow\, \rangle + | \downarrow\, \uparrow\, \rangle,
| \downarrow\, \downarrow\, \rangle,
| \uparrow\, \downarrow\, \rangle - | \downarrow\, \uparrow\, \rangle \Big\}
,
\end{equation}
such that the diagonalized Hamiltonian becomes
\begin{equation}\label{L2eigenvaluesalt}
%{\rm \bf{H}}=
\begin{pmatrix}
0 & 0 & 0 & 0 \\
0 & 0 & 0 & 0 \\
0 & 0 & 0 & 0 \\
0 & 0 & 0 & 8 \\
\end{pmatrix}
.
\end{equation}
In view of \eqref{schroedinger} we see that we have two distinct energy eigenvalues, namely a triply degenerate value $E=0$ with eigenstates $| \uparrow\, \uparrow\, \rangle, | \uparrow\, \downarrow\, \rangle + | \downarrow\, \uparrow\, \rangle,
| \downarrow\, \downarrow\, \rangle$, as well as a non-degenerate eigenvalue $E=8$ with eigenstate $| \uparrow\, \downarrow\, \rangle - | \downarrow\, \uparrow\, \rangle$. Group theoretically this is to be expected, as we have a $\alg{su}(2)$ invariant chain, and in the tensor product of two (since $L=2$) spin-$\frac{1}{2}$ representations we have one spin-1 triplet and one spin-0 singlet: $\frac{1}{2} \otimes \frac{1}{2} = 1 \oplus 0$. Recall that the Hamiltonian {\bf H} commutes with the total spin, therefore, as remarked below \eqref{periodic}, the energy eigenvalues in each $\alg{su}(2)$ multiplet must be identical.

Let us introduce two further important operators. The first is the {\it permutation operator} $\mathbb{P}_{l,l+1}$, which permutes the spins at site $l$ and $l+1$. In our simplest length $L=2$ example, we clearly have, in the basis \eqref{L2basis},
\begin{equation}\label{permutation}
\mathbb{P}_{1,2}=
\begin{pmatrix}
1 & 0 & 0 & 0 \\
0 & 0 & 1 & 0 \\
0 & 1 & 0 & 0 \\
0 & 0 & 0 & 1 \\
\end{pmatrix}
.
\end{equation}
If we denote by $\mathbb{I}_{1,2}$ the four-dimensional unit matrix, we can obviously rewrite the Hamiltonian matrix \eqref{hamiltonianL2} as ${\rm \bf{H}}=2\,(\mathbb{I}_{1,2}-\mathbb{P}_{1,2})+2\,(\mathbb{I}_{2,1}-\mathbb{P}_{2,1})$. Since the interaction in the Hamiltonian \eqref{hamiltonian} for general $L$ is nearest-neighbor and pairwise, we can immediately lift this result from 2 to $L$ and find that
\begin{equation}\label{hamiltonian2}
{\rm \bf{H}}=2\, \sum_{l=1}^{L}\left(\mathbb{I}_{l,l+1}-\mathbb{P}_{l,l+1}\right)
.
\end{equation}
It is a nice exercise with tensor products to alternatively deduce this result directly from \eqref{hamiltonian} by using the explicit form of the three Pauli matrices. For the solution of this exercise, see again \cite{Nepomechie:1998jf}.

The other important operator is the {\it shift operator}, defined as ${\rm \bf{U}}=\mathbb{P}_{1,2}\,\ldots\,\mathbb{P}_{L-1,L}$. We invite the reader to playfully act with it on a general spin chain state, and its name will immediately become obvious. Strictly speaking, we could define a left-shift and a right-shift operator, but this distinction will not be needed here. Now it is easy to see that because of the periodic boundary conditions the shift operator commutes with the Hamiltonian: $[{\rm \bf{U}},{\rm \bf{H}}]=0$, as well as with the total spin operator:  $[{\rm \bf{U}},\vec S]=0$. Therefore each eigenstate $|\psi\rangle$ in \eqref{schroedinger} must also carry a definite shift eigenvalue $U$: ${\rm \bf{U}} \cdot |\psi\rangle=U\,|\psi\rangle$. Furthermore, it should be obvious that ${\rm \bf{U}}^L=\mathbb{I}$, since shifting by $L$ sites returns us to the original configuration. Therefore, all eigenvalues of ${\rm \bf{U}}$ have to satisfy $U^L=1$, and we conclude that they are quantized in units of $1/L$, i.e. $U=e^{2\pi i n/L}$ with the quantum number $n=0,1,\ldots,L-1$. In our nearly trivial $L=2$ example we have ${\rm \bf{U}}=\mathbb{P}_{1,2}$ as in \eqref{permutation}, while the shift operator becomes diagonal in the basis \eqref{L2beigenbasisalt} where it reads
\begin{equation}\label{shiftevs}
%{\rm \bf{H}}=
\begin{pmatrix}
+1 & 0 & 0 & 0 \\
0 & +1 & 0 & 0 \\
0 & 0 & +1 & 0 \\
0 & 0 & 0 & -1 \\
\end{pmatrix}
.
\end{equation}
So the triplet state has shift quantum number $n=0$, and the singlet $n=1$. The shift operator plays a crucial role in the ${\cal N}=4$  gauge theory context: Because of gauge invariance, all planar local composite operators are single trace operators, see part I of this review. The trace leads to {\it two distinct} consequences for the spin chain interpretation of these operators. The {\it first} is that the ``first'' field (spin chain site $l=1$) inside the trace gets ``hooked'' to the ``last field'' (site $l=L$). This is just the periodic boundary condition \eqref{periodic}. The {\it second} is that because of trace cyclicity only the eigenvalue $U=1$ is allowed, all states with $n\neq0$ are actually identically zero in gauge theory. So if we interpret our $L=2$ eigenstates in \eqref{L2beigenbasisalt} via \eqref{singlespin} in gauge theory, we find, using trace cyclicity, 
\begin{equation}\label{L2gauge}
\Big\{ 
\Tr Z^2,
\Tr Z X + \Tr X Z,
\Tr X^2,
\Tr Z X - \Tr X Z
\Big\}
=
\Big\{ 
\Tr Z^2,
2\, \Tr Z X,
\Tr X^2,
0
\Big\}
.
\end{equation}
So as concerns the triplet, $\Tr Z X$ and $\Tr X^2$ are some $\alg{su}(2)$ descendents of the BPS primary state $\Tr Z^2$ (cf.~part Va) with anomalous dimension $E=0$, but the $L=2$ Heisenberg chain's singlet simply disappears (or maybe better to say: is projected out) in gauge theory! So the energy $E=8$ of this state has no interpretation in gauge theory, at least not in the maximally supersymmetric ${\cal N}=4$ case.

\subsection{Understanding the solution of the problem}

As we just explained, the Hamiltonian is just a $2^L \times 2^L$ matrix, so what is the problem? The quick answer is that Hans Bethe neither had a laptop nor Mathematica or Maple (and even if he had had one, he would have quickly run into problems for sizes $L \ge 10$ or so). So he derived the equations which carry his name. Let us assume that we have a number $M$ of down spins in a chain of length $L$. While the Hamiltonian clearly shifts around those down spins, it certainly does not change $M$, as is easily seen from \eqref{hamiltonian2}. In fact, this follows from \eqref{totalspin}, which says, in particular, that the z-component of spin commutes with {\bf H}: $[{\rm {\bf H}},S^z]=0$: For all-spins-up, the eigenvalue of $S^z$ is $\frac{1}{2} L$, and reversing one spin lowers the total spin by 1 ({\it not} by $\frac{1}{2}$, think about it!) so we have for $M$ reversed spins $\frac{1}{2}(L-2 M)$. Since $S^z$ and $L$ are conserved, so is $M$. So the $2^L \times 2^L$ matrix is block-diagonal, with $L+1$ blocks ($M=0,\ldots,L$), and the $M$-th block is a $\binom{L}{M} \times \binom{L}{M} $ matrix. In order to nicely write down the eigenvalues of the $M$-th block, Bethe introduced $M$ complex numbers $u_1,\ldots,u_M$. To be honest, he probably first introduced them as real numbers, but quickly found that they needed to be complex for his solution to work in general. According to him, the eigenvalues of the Hamiltonian {\bf H} (the energies) are then given by 
\begin{equation}\label{energy}
E=2\,\sum_{k=1}^{M}\frac{1}{u_{k}^{2}+\frac{1}{4}}\, ,
\end{equation}
and the eigenvalues of the shift operator {\bf U} are
\begin{equation}\label{shift}
U=
\prod_{k=1}^M \frac{u_k+\frac{i}{2}}{u_k-\frac{i}{2}}
\, .
\end{equation}
So, instead of diagonalizing a $\binom{L}{M} \times \binom{L}{M}$ matrix, we have to find the correct sets of {\it distinct}\footnote{%
We will see in the next subsection why the Bethe roots must all be distinct.
}
numbers $\{u_1,\ldots,u_M\}$. They are fittingly called Bethe roots, and are determined from a system of $M$ algebraic equations for these $M$ variables:
\begin{equation}\label{betheeqsnotwist}
\Big( \frac{u_k+\frac{i}{2}}{u_k-\frac{i}{2}}\Big) ^L =\prod_{\stackrel{j=1}{j\neq k}}^{M} \frac{u_k-u_j+i}{u_k-u_j-i}\, ,
\quad {\rm where} \quad
k=1,\ldots,M
\, .
\end{equation}
So what are the solutions of these equations? Let us see how to reproduce our results for the $L=2$ chain from the previous section. There are three blocks of {\bf H}, corresponding to $M=0,1,2$ (look back at \eqref{hamiltonianL2}). The $M=0$ block is $1 \times 1$, there are no Bethe roots, so the sum in \eqref{energy} is empty and indeed gives $E=0$. The $M=1$ block is of size $2 \times 2$. Solving \eqref{betheeqsnotwist}, there is just one {\it finite} Bethe root, which is easily found to be $u_1=0$. This indeed gives from \eqref{energy} E=8 and from \eqref{shift} U=-1. The reader can also try the final $1 \times 1$ block with $M=2$, she will, however, not find finite, distinct Bethe roots $u_1, u_2$ corresponding to the eigenvalue $E=0$ for $|\downarrow \downarrow \rangle$. 

This was a partial success, we did encounter the correct energies $E=0,8$ which appear at $L=2$, but the multiplicities do not seem right. Why did we only find two instead of four states for the $L=2$ chain? The answer is pretty tricky, and best understood by manipulating the periodic boundary conditions \eqref{periodic}, replacing them by 
\begin{equation}\label{twisted}
S^3_{L+1}:= S^3_1\, ,
\qquad \qquad
S^{\pm}_{L+1}:=e^{\mp i \phi}\, S^{\pm}_1
\, ,
\end{equation}
%
%\begin{equation}\label{hamiltoniantwist}
%{\bf H}_\phi=4\, \sum_{l=1}^{L}\big[ \frac{1}{4}-S_{l}^{3}\,S_{l+1}^{3}-\frac{1}{2}\,e^{+i\frac{\phi}{L}}\,S_{l}^{+}\,S_{l+1}^{-}-\frac{1}{2}\,e^{-i%\frac{\phi}{L}}\,S_{l}^{-}\,S_{l+1}^{+}\big]\, ,
%\end{equation}
%
where $\phi$ is some phase, and $S_l^{\pm}=S_l^1\pm i\, S_l^2$ are the usual $\alg{su}(2)$ ladder operators. For $\phi=0$ we obviously recover \eqref{periodic}. Let us denote \eqref{hamiltonian} with \eqref{twisted} instead of \eqref{periodic} by ${\rm \bf{H}}_\phi$. The length $L=2$ Hamiltonian matrix \eqref{hamiltonianL2} is then modified to 
\begin{equation}\label{hamiltonianL2twist}
{\rm \bf{H}}_\phi=
\begin{pmatrix}
0 & 0 & 0 & 0 \\
0 & +2 & -2& 0 \\
0 & -2 & +2 & 0 \\
0 & 0 & 0 & 0 \\
\end{pmatrix}
+
\begin{pmatrix}
0 & 0 & 0 & 0 \\
0 & +2 & -2\,e^{+i\,\phi}& 0 \\
0 & -2\,e^{-i\,\phi} & +2 & 0 \\
0 & 0 & 0 & 0 \\
\end{pmatrix}
.
\end{equation}
%
%
%\begin{equation}%\label{hamiltonianL2twist}
%{\rm \bf{H}}_\phi=
%\begin{pmatrix}
%0 & 0 & 0 & 0 \\
%0 & 2 & -2\,e^{+i\frac{\phi}{2}}& 0 \\
%0 & -2\,e^{-i\frac{\phi}{2}} & 2 & 0 \\
%0 & 0 & 0 & 0 \\
%\end{pmatrix}
%+
%\begin{pmatrix}
%0 & 0 & 0 & 0 \\
%0 & 2 & -2\,e^{-i\frac{\phi}{2}}& 0 \\
%0 & -2\,e^{+i\frac{\phi}{2}} & 2 & 0 \\
%0 & 0 & 0 & 0 \\
%\end{pmatrix}
%.
%\end{equation}
%
The basis \eqref{L2beigenbasisalt} is modified to
\begin{equation}\label{L2beigenbasisalttwist}
\Big\{ | \uparrow\, \uparrow\, \rangle,
e^{-i\frac{\phi}{4}}| \uparrow\, \downarrow\, \rangle + e^{+i\frac{\phi}{4}}| \downarrow\, \uparrow\, \rangle,
| \downarrow\, \downarrow\, \rangle,
e^{-i\frac{\phi}{4}} | \uparrow\, \downarrow\, \rangle - e^{+i\frac{\phi}{4}} | \downarrow\, \uparrow\, \rangle \Big\}
,
\end{equation}
and the diagonalized Hamiltonian now reads in generalization of \eqref{L2eigenvaluesalt}
\begin{equation}\label{L2eigenvaluesalttwist}
%{\rm \bf{H}}=
\begin{pmatrix}
0 & 0 & 0 & 0 \\
0 & 8\,\sin^2 \frac{\phi}{4} & 0 & 0 \\
0 & 0 & 0 & 0 \\
0 & 0 & 0 & 8\,\cos^2 \frac{\phi}{4}\\
\end{pmatrix}
.
\end{equation}
We see that the degeneracy of the triplet is (partially) lifted, as the 
%state $| \uparrow \downarrow\, \rangle + | \downarrow\, \uparrow\, \rangle$ 
``middle state''
now has energy $E=8\,\sin^2 \frac{\phi}{4}$ while $| \uparrow\, \uparrow\, \rangle$ and $| \downarrow\, \downarrow\, \rangle$ remain at $E=0$. Therefore the $\alg{su}(2)$ invariance must be broken\footnote{%
We still have $[{\rm {\bf H}}_\phi,S^z]=0$, so the number $M$ of down-spins is still conserved.
}
for generic $\phi$, and we indeed now have $[ {\rm \bf{H}}_\phi, \vec S] \neq 0$.

Bethe's equations still work with minor modifications. In fact, they work much better! The formula for the energy \eqref{energy} remains unaffected, but \eqref{betheeqsnotwist} are modified to
\begin{equation}\label{betheeqs}
\Big( \frac{u_k+\frac{i}{2}}{u_k-\frac{i}{2}}\Big) ^L e^{i\phi}=\prod_{\stackrel{j=1}{j\neq k}}^{M} \frac{u_k-u_j+i}{u_k-u_j-i}\, ,
\quad {\rm where} \quad
k=1,\ldots,M
\, .
\end{equation}
Let's redo the $L=2$ case (confer the discussion just after \eqref{betheeqsnotwist}).
There are still three blocks of {\bf H}, corresponding to $M=0,1,2$. As before, the $M=0$ block is $1 \times 1$, there are no Bethe roots, and thus $E=0$. The $M=1$ block is of size $2 \times 2$. However, this time around, solving \eqref{betheeqs} yields not one but {\it two}  finite Bethe roots. They are easily found to be $u_1=-\frac{1}{2} \cot \frac{\phi}{4}$ or $u_1=\frac{1}{2} \tan \frac{\phi}{4}$. From \eqref{energy} this gives, respectively,  $E=8\,\sin^2 \frac{\phi}{4}$ and $E=8\,\cos^2 \frac{\phi}{4}$. Finally, for the $1 \times 1$ block with $M=2$  we find $u_{1,2}=-\frac{1}{2} \cot \frac{\phi}{2}\pm \frac{i}{2 \sin\frac{\phi}{2}} $. Curiously, this leads from \eqref{energy} for arbitrary $\phi$ to the correct energy $E=0$, showing that ``non-trivial'' Bethe roots can lead to trivial energies. Let us now take the $\phi \rightarrow 0$ limit. We observe that the Bethe root $u_1=-\frac{1}{2} \cot \frac{\phi}{4}$ of the $M=1$ descendent $| \uparrow \downarrow\, \rangle + | \downarrow\, \uparrow\, \rangle$ of the $\alg{su}(2)$ highest weight state $|\uparrow \uparrow\rangle$ {\it diverges}, while the root $u_1=\frac{1}{2} \tan \frac{\phi}{4}$ of the singlet state $| \uparrow \downarrow\, \rangle - | \downarrow\, \uparrow\, \rangle$ returns to its correct untwisted value $u_1=0$. Likewise, the two Bethe roots $u_{1,2}=-\frac{1}{2} \cot \frac{\phi}{2}\pm \frac{i}{2 \sin\frac{\phi}{2}} $ of the $M=2$ descendent $|\downarrow \downarrow \rangle$ also both diverge. So this is the answer to our multiplicity puzzle: The untwisted Bethe ansatz equations \eqref{betheeqsnotwist} only yield the highest weight states,\footnote{
It may be proved that the eigenvectors are highest weight states and,  therefore, in \eqref{betheeqsnotwist} $M$ should in fact be restricted to values $M \le L/2$.
} 
i.e.~those states which are annihilated by the total $S^+$.
 The descendents of these formally correspond to adding roots at infinity! Note that each ``step'' down the multiplet adds one further such infinite root.\footnote{%
Note that if we allow roots at infinity, we have to {\it ad hoc} relax the restriction that all Bethe roots must be distinct, since we need $n$ of these for level $n$ descendents.
}
This is an artifact resulting from the $\alg{su}(2)$ invariance, if we break the latter by the twist field $\phi$, the Bethe equations \eqref{betheeqs} yield the correct energies and all states are nicely described by finite sets of Bethe roots.

The attentive reader might wonder what is the advantage of this reformulation of the matrix diagonalization problem \eqref{schroedinger} by \eqref{betheeqsnotwist} with \eqref{energy} (or their generalization \eqref{betheeqs} with \eqref{energy}). The system of equations \eqref{betheeqs} is increasingly tricky to solve, even using numerical techniques, as $L$ and $M$ increase from the nearly trivial values we just discussed. Nevertheless there are huge advantages as compared to brute force diagonalization of the $2^L \times 2^L$ Hamiltonian. In fact, in order to understand this statement, we invite the serious reader to take Mathematica or Maple, and to find the complete spectrum of the $L=3,4$ and maybe the $L=5$ chain by finding all states both from \eqref{betheeqs} as well as directly from \eqref{hamiltonian} with ``quasiperiodic'' boundary conditions \eqref{twisted}. For sure, as $L$ increases, direct diagonalization becomes impossible even with the help of a powerful computer due to the exponential growth of the matrix size. With a ``metal'' where the number of atoms in a unit volume is $L^3\sim{\cal O}(10^{23})$ this is clearly impossible. On the other hand, the system of equations \eqref{betheeqs} actually tends to enormously simplify for large values of $L$: One often is able to derive neat linear integral equations in this thermodynamic limit. For one example within this review series, see \cite{chapCurve}.
Furthermore, the reformulation of a matrix diagonalization problem to an entirely algebraic problem is conceptually very interesting and useful. Note that this algebraic reformulation is quite different from working out the characteristic polynomial $\det (E\, \mathbb{I}-\bf{H})$ of the Hamiltonian matrix, where we first need to compute a large $2^L \times 2^L$ determinant. In fact, numerically it is a particularly bad idea to compute the characteristic polynomial and to then determine its eigenvalues, since it  wildly oscillates. It is much better to obtain the spectrum by different methods starting from the original matrix.

\subsection{Understanding how to arrive at the solution, and AdS/CFT}

So far we just tried to explain how Bethe's solution of the diagonalization problem of the Heisenberg chain works. But how to find his equations \eqref{energy}, \eqref{betheeqs}? If an exact solution to some problem exists, there are usually many ways to find it. The Heisenberg chain is no exception. In the following, we will briefly sketch a number of rather distinct, interesting solution methods, referring for details to various excellent pedagogical presentations already existing in the literature. We will begin in Section \ref{sec:CBA} with Bethe's original method, nowadays called ``coordinate Bethe ansatz''. We then move on to a more ``modern'' approach termed ``algebraic Bethe ansatz'' in Section \ref{sec:ABA}. In Section \ref{sec:AdS/CFT} we briefly discuss how these techniques are related to AdS/CFT, as well as to other articles in this review collection. 
%We also briefly scan through a number of interesting generalizations of Bethe's original equations, and discuss which of these might be useful for gauge-string dualities. 
We end in Section \ref{sec:Baxter} with yet another method of solution pioneered a long time ago by Baxter. Curiously, it was only very recently properly applied to the Heisenberg magnet. This author believes that this method will prove to be very powerful in the AdS/CFT context.

%%%%%%%%%%%%%%%%%%%%%%%%%%%%%%%%%%%%%%%%%%%%%%%%%%%%%%%%%%%%%%%%%%%%%%%%%%%%%%%%
\section{Coordinate Bethe Ansatz}
\label{sec:CBA}

``Ansatz'' is a German word for a procedure which means ``make a guess for the solution, and check whether it works''. Bethe made an inspired guess for the form of the eigenvector $|\psi \rangle$ in \eqref{schroedinger}, and then constructively proved that his ansatz is correct if certain conditions, the {\it Bethe equations}, are satisfied. As a by-product, the energy eigenvalues $E$ are found.

Clearly $|\psi \rangle$ may be written for a given number of down spins $M$ as
\begin{equation}\label{wvfnctn}
|\psi \rangle
=
\sum \psi(l_1,l_2,\ldots,l_M)\,S^-_{l_1}\,S^-_{l_2}\,\dots\,S^-_{l_M}\,|0\rangle\, ,
\end{equation}
where $|0\rangle=|\uparrow \uparrow \ldots \uparrow \rangle$ is the {\it vacuum} state where all $L$ spins point up, and the (local) $\alg{su}(2)$ lowering operator $S^-_{l_k}$ flips the spin from up to down at position $l_k$. We can think of $\psi(l_1,l_2,\ldots,l_M)$ as the position space wave function of the spin chain, where the positions $l_k$ live on the lattice numbered by $1,\ldots,L$. The sum in \eqref{wvfnctn} is over all orderings $1\leq l_1<l_2<\,\ldots\,l_M\leq L$, in order to avoid overcounting of states. The $<$ stems from the fact that we can only lower each up-spin once, since each lattice site is in a spin $\frac{1}{2}$ representation. Of course we could have just as well started from $|0\rangle=|\downarrow \downarrow \ldots \downarrow \rangle$, and then used $S^+_{l_k}$ instead of $S^-_{l_k}$ in \eqref{wvfnctn}. But as in real life, one is often forced to make a choice in order to proceed.

So far so good, there is nothing ``Bethe'' yet. Here is his ansatz:
\begin{equation}\label{cba}
\psi(l_1,l_2,\ldots,l_M)
=
\sum_{\{\tau\}} A(\tau)\,e^{i\,p_{\tau_1}\,l_1 + \ldots + i\,p_{\tau_M}\,l_M}
\, .
\end{equation}
The sum runs over the set $\{\tau\}$ of all $M!$ permutations $\tau$ of the $M$ downspins, so $\tau=\{\tau_1,\ldots,\tau_M\}$ is a permutation of the $M$ labels $1,\ldots,M$. This looks like a clever linear superposition of $M!$ plane wave factors, where each factor is multiplied with an amplitude $A(\tau)$ dependent on the permutations $\tau$, but {\it not} on the positions $l_k$. We can also think of this as a kind of generalization of a Fourier transform, which usually solves translation-invariant free systems. Our system is not free, however! This picture is nevertheless useful, as it leads to the interpretation of the set of numbers $\{p_1,\ldots,p_M\}$ as the lattice {\it momenta} of the $M$ down-spins in the background of the up-spin vacuum.

The next step is to check whether \eqref{cba} really works. So, with some effort we can just plug this expression into \eqref{wvfnctn}, and check that we indeed have a solution (i.e.~\eqref{schroedinger} holds) iff the momenta $\{p_1,\ldots,p_M\}$ satisfy for $k=1,\ldots,M$ the constraints\footnote{%
It is technically easier to first try $M=1$, which trivially works, then $M=2$, where we find the nice formula for $S(p_k,p_j)$, and then go about proving it for general $M$.
}
\begin{equation}\label{cbaeqs}
e^{i\,p_k\,L} e^{i\,\phi}
=
\prod_{\stackrel{j=1}{j\neq k}}^{M} S(p_k,p_j)
\, ,
\quad {\rm where} \quad
S(p_k,p_j)=-\frac{e^{i\,p_k+i\,p_j}-2\,e^{i\,p_k}+1}{e^{i\,p_k+i\,p_j}-2\,e^{i\,p_j}+1}
\, .
\end{equation}
In this case, the amplitudes $A(\tau)$ in the wavefunction \eqref{cba} are given by
\begin{equation}\label{amplitudes}
A(\tau)
=
{\rm sign}(\tau)\,\prod_{j<k}
\left( e^{i\,p_k+i\,p_j}-2\,e^{i\,p_k}+1\right)
,
\end{equation}
where ${\rm sign}(\tau)$ is the signature of the permutation, 
while the energy eigenvalue is found to be
\begin{equation}\label{energycba}
E
=
\sum_{k=1}^M 8\,\sin^2 \left(\frac{p_k}{2}\right)
.
\end{equation}
%
%Finally it is also easy to derive the shift eigenvalue, which is in line with intuition
%
%\begin{equation}\label{shiftcba}
%U
%=
%\prod_{k=1}^M e^{i\,p_k}
%\, .
%\end{equation}
%
%As an exercise, we invite the reader to generalize these formulas to the case $\phi \neq 0$, using the boundary conditions \eqref{twisted}.

Comparing, respectively, the expressions \eqref{energycba} and \eqref{cbaeqs} to \eqref{energy} and \eqref{betheeqsnotwist}, even the hasty reader will recognize their similarity. These are the same equations, once we identify for all $k=1,\ldots,M$
\begin{equation}\label{raptomom}
e^{i p_k}=\frac{u_k+\frac{i}{2}}{u_k-\frac{i}{2}}\ \ \  \Longleftrightarrow \ \ \ u_k=\frac{1}{2}\cot \frac{p_k}{2}\, .
\end{equation}
So the Bethe roots are nothing but specially parametrized lattice momenta of the down-spin ``particles'', which are often called magnons.

The nice thing about the Bethe ansatz is that it not only yields the spectrum, but also the (unnormalized) wavefunctions\footnote{%
The normalization may also be found, see e.g.~\cite{Korepin:1993aa}.
}.
It is easy to see that the latter are fully antisymmetric under exchange of any two momenta and, therefore, any two Bethe roots. This is the reason, mentioned already just before \eqref{betheeqsnotwist}, why we need to discard solutions with coinciding roots, as in this case the eigenvector $|\psi\rangle$ of \eqref{schroedinger} vanishes, which is of course disallowed by elementary linear algebra.

We recommend to the serious student to study the solution we just sketched in much more detail. Apart from Bethe's quite readable original paper \cite{Bethe:1931aa}, a very pedagogical presentation of the coordinate Bethe ansatz may be found in \cite{Karbach:1998aa}. Chapter 2.1 of \cite{Staudacher:2004tk} might also be helpful. Insightful and artfully written accounts by B.~Sutherland are found in \cite{Sutherland:1978aa} and in his book \cite{Sutherland:2004aa}.

%%%%%%%%%%%%%%%%%%%%%%%%%%%%%%%%%%%%%%%%%%%%%%%%%%%%%%%%%%%%%%%%%%%%%%%%%%%%%%%%
\section{Algebraic Bethe Ansatz}
\label{sec:ABA}

The coordinate Bethe ansatz is very ``physical'', and widely applicable. However, one disadvantage is that it totally obscures {\it why} a given Hamiltonian is integrable. A beautiful general method originates in work of Baxter in the early 1970's, and was systematized and generalized in the late 1970's and early 1980's within the so-called ``quantum inverse scattering program'' initiated by the ``Leningrad school'' around Ludvig Faddeev. Its main advantage is that it allows to find in a rather systematic way very general classes of integrable models. For example, it is easy to generalize the XXX Heisenberg Hamiltonian to more general representations of the spin, and to symmetry algebras larger than $\alg{su}(2)$ while preserving integrability.

Let us panoramically sketch its key features. I certainly cannot improve the brilliant presentation in \cite{Faddeev:1996iy}. If this is too hard upon initial reading, please first study \cite{Nepomechie:1998jf}. Some important complementary information is in \cite{Faddeev:1994nk}, and the presentation in the very recent notes \cite{Doikou:2009aa} as well as in the initial review part of the article \cite{Escobedo:2010xs} is also very nice.

The starting point is not the Hamiltonian \eqref{hamiltonian}, but instead a ``generating object'' \cite{Faddeev:1996iy}, the quantum Lax operator. It admittedly falls a bit from the sky; I do not know a very good way to motivate it. Then again, one has to start with {\it something}\footnote{%
In the case of AdS/CFT, we neither know the Hamiltonian nor a good ``generating object''.
}
(mathematicians call it axiom). In the case of the Heisenberg chain, this operator reads
\begin{equation}\label{laxfaddeev}
\mathcal{L}_{a,l}(u)=
\left( \begin{array}{cc}
u+ i\,S_l^{3} &  ~~i\,S_l^- \\
i\,S_l^+ & ~~u-i\,S_l^{3}  \end{array} \right)_a,
\end{equation}
which is a $2 \times 2$ matrix in some {\it auxiliary space} $\mathbb{C}^2$ indexed by $a$. Each of its four matrix elements is also a $2 \times 2$ matrix expressed through the site-$l$ spin operators we introduced in \eqref{hamiltonian}. It also depends on a complex variable $u$, the {\it spectral parameter}. A {\it monodromy matrix} is then built as\footnote{%
Here $\cdot$ denotes $2 \times 2$ matrix multiplication in the auxiliary space. The entries of this $2 \times 2$ matrix act on \eqref{quantumspace}. Therefore, the whole thing $\mathcal{M}_a(u)$ acts on the tensor product of \eqref{quantumspace} and the auxiliary space $a$.
}
\begin{equation}\label{monodromy}
\mathcal{M}_a(u)=
\left( \begin{array}{cc}
e^{i\,\frac{\phi}{2}} & 0 \\
0  & e^{-i\,\frac{\phi}{2}}  \end{array} \right)\cdot
\mathcal{L}_{a,L}(u)\cdot\mathcal{L}_{a,L-1}(u)\cdot \ldots \cdot \mathcal{L}_{a,2}(u)\cdot\mathcal{L}_{a,1}(u)
\, .
\end{equation}
One next takes the trace $\Tr_a$ over the two-dimensional auxiliary space $a$
\begin{equation}\label{transfertrace}
\mathbf{T}(u)=\Tr_a\, \mathcal{M}_a(u)
\, ,
\end{equation}
and thereby constructs the {\it transfer matrix} as an operator on the quantum space \eqref{quantumspace}, which also depends on $u$. Now take {\it two} different auxiliary spaces $a$ and $b$ instead of just one, while concentrating on a single spin chain site $l$.
%
%\begin{equation}\label{laxfaddeevtwice}
%\mathcal{L}_{a,l}(u)=
%\left( \begin{array}{cc}
%u+ i\,S_l^{3} &  ~~i\,S_l^- \\
%i\,S_l^+ & ~~u-i\,S_l^{3}  \end{array} \right)_a,
%\quad {\rm and} \quad
%\mathcal{L}_{b,l}(u)=
%\left( \begin{array}{cc}
%u+ i\,S_l^{3} &  ~~i\,S_l^- \\
%i\,S_l^+ & ~~u-i\,S_l^{3}  \end{array} \right)_b,
%\end{equation}
%
%where the left $2 \times 2$ matrix acts on space $a$, and the right $2 \times 2$ matrix on $b$. 
Then you can (and should, at least once in your life) check by direct computation that the {\it Yang-Baxter equation} holds on the triple tensor product of our three spaces $a,b,l$:
\begin{equation}\label{ybe}
\mathcal{R}_{a,b}(u-u')\,\mathcal{L}_{a,l}(u)\,\mathcal{L}_{b,l}(u')=\mathcal{L}_{b,l}(u')\,\mathcal{L}_{a,l}(u)\,\mathcal{R}_{a,b}(u-u').
\end{equation}
The {\it R-matrix} $\mathcal{R}$ is essentially, in hopefully obvious notation, the same thing as \eqref{laxfaddeev}
\begin{equation}\label{R-matrix}
\mathcal{R}_{a,b}(u)=
\left( \begin{array}{cc}
u+\frac{i}{2}+i\,S_a^{3} &  ~~i\,S_a^- \\
i\,S_a^+ & ~~u+\frac{i}{2}-i\,S_a^{3}  \end{array} \right)_b
=
\left( \begin{array}{cc}
u+\frac{i}{2}+i\,S_b^{3} &  ~~i\,S_b^- \\
i\,S_b^+ & ~~u+\frac{i}{2}-i\,S_b^{3}  \end{array} \right)_a.
\end{equation}
Using the notation of Section \ref{sec:Intro}, there are further instructive ways to write this
\begin{equation}\label{R-matrix2}
\mathcal{R}_{a,b}(u)
=
\left(u+\frac{i}{2}\right)\,\mathbb{I}_{a,b}+2\,i\,S^3_a\,S^3_b+i\,S^+_a\,S^-_b+i\,S^-_a\,S^+_b
=
u+\frac{i}{2}+2\,i\,\vec S_a\cdot\vec S_b
\, ,
%u\, \mathbb{I}_{a,b}+i\, \mathbb{P}_{a,b}
\end{equation}
the most important form being
\begin{equation}\label{R-matrix3}
\mathcal{R}_{a,b}(u)
=
u\, \mathbb{I}_{a,b}+i\, \mathbb{P}_{a,b}\, .
\end{equation}
You should also learn the beautiful graphical way (the so-called  ``train arguments'') to depict \eqref{ybe}, which allows to trivialize many proofs (this is one of the things very nicely explained in the earlier Faddeev lecture \cite{Faddeev:1994nk}). For example this one,
\begin{equation}\label{rmm}
\mathcal{R}_{a,b}(u-u')\,\mathcal{M}_a(u)\,\mathcal{M}_b(u')=\mathcal{M}_b(u')\,\mathcal{M}_a(u)\,\mathcal{R}_{a,b}(u-u'),
\end{equation}
where $\,\mathcal{M}_a$ and $\,\mathcal{M}_b$ are built as in \eqref{monodromy}, using \eqref{laxfaddeev}. Then taking the doubletrace $\Tr_a\,\Tr_b$ of \eqref{rmm} over the two auxiliary spaces $a,b$ the matrix $\mathcal{R}_{a,b}$ drops out, and we derive
\begin{equation}\label{TT}
\mathbf{T}(u)\,\mathbf{T}(u')= \mathbf{T}(u')\,\mathbf{T}(u),
\quad {\rm i.e.} \quad 
\left[\mathbf{T}(u),\mathbf{T}(u')\right]=0\, .
\end{equation}
The transfer matrix operator commutes with itself at different values of the spectral parameter $u$!
What does all this formal stuff have to do with our earlier discussion, though? The point is that we can expand $\mathbf{T}(u)$, or actually more naturally $\log \mathbf{T}(u)$ in a power series around any point $u_0$ of the complex $u$-plane, thereby generating a set of 
linearly independent operators acting on the quantum space \eqref{quantumspace}. Because of \eqref{TT}, these all commute with each other (or, to express this in fancier way, ``are in involution''). This formally proves the integrability, since for the special point $u_0=\frac{i}{2}$ one of these charges is our Hamiltonian \eqref{hamiltonian} with boundary conditions \eqref{twisted}:
\begin{equation}\label{HfromT}
{\rm \bf{H}}_\phi
=2\,L-2\,i\,\frac{d}{d u} \log \mathbf{T}(u) \Big|_{u=\frac{i}{2}}
\, .
\end{equation}
What is more, we may also obtain the Bethe equations within this formalism, even though it is somewhat tedious (see chapter 4 of \cite{Faddeev:1996iy}). The basic idea is quite nice and simple though. First, introduce the following notation for the monodromy \eqref{monodromy}
\begin{equation}\label{abcd}
\mathcal{M}_a(u)=
\left( \begin{array}{cc}
\mathbf{A}(u) & \mathbf{B}(u) \\
\mathbf{C}(u)  & \mathbf{D}(u)  \end{array} \right)
\, ,
\end{equation}
where again the $2 \times 2$ matrix acts on the auxiliary space $a$, and in consequence $\mathbf{A},\mathbf{B},\mathbf{C},\mathbf{D}$ are operators on \eqref{quantumspace}. If one is just interested in the transfer matrix operator $\mathbf{T}(u)=\mathbf{A}(u)+\mathbf{D}(u)$ the off-diagonal components $\mathbf{C}(u)$ and $\mathbf{B}(u)$ may be ignored. They are, however, very useful for the construction of the Bethe states. Let us use the same ``ferromagnetic'' vacuum state $|0\rangle=|\uparrow \uparrow \ldots \uparrow \rangle$ as for the coordinate Bethe ansatz, c.f.~\eqref{wvfnctn}. One immediately sees from \eqref{laxfaddeev},\eqref{monodromy} that $\mathbf{C}(u) |0\rangle=0$. Now the algebraic Bethe ansatz is simply the following ``trial wavefunction''
\begin{equation}\label{aba}
|\psi \rangle
=
\mathbf{B}(u_1)\,\mathbf{B}(u_2)\,\dots\,\mathbf{B}(u_M)\,|0\rangle\, ,
\end{equation}
which should be compared to \eqref{wvfnctn},\eqref{cba}. The next step is again to check whether \eqref{aba} really works. What does that mean here? Well, this time around we have to check whether the state \eqref{aba} is an eigenstate of the transfer matrix {\it operator} $\mathbf{T}(u)$, i.e.~that
\begin{equation}\label{transferev}
 \mathbf{T}(u) \cdot |\psi\rangle=T(u)\,|\psi\rangle\, ,
\end{equation}
where $T(u)$ is the eigenvalue. Then we will have killed (=diagonalized) $L$ birds (=commuting charges) with one stone, and thus will have  also solved \eqref{schroedinger} (because of \eqref{HfromT}). The tedious part is to check this, where one proceeds again from \eqref{rmm}. The upshot is that it actually does not work (i.e.~the state $|\psi \rangle$ in \eqref{aba} is not an eigenstate) {\it unless} the set of variables $\{u_1,u_2,\ldots,u_M\}$ in \eqref{aba} satisfies the Bethe equations \eqref{betheeqs}. In fact, they follow from eliminating the ``unwanted terms'' ruining \eqref{transferev}. Pretty cool!

%%%%%%%%%%%%%%%%%%%%%%%%%%%%%%%%%%%%%%%%%%%%%%%%%%%%%%%%%%%%%%%%%%%%%%%%%%%%%%%%
\section{Extensions, Deformations, and AdS/CFT}
\label{sec:AdS/CFT}

We just sketched two methods (coordinate and algebraic Bethe ans\"atze) for solving the one-dimensional Heisenberg magnet. The latter reemerged some 80 years after its invention in the special $\alg{su}(2)$ sector of AdS/CFT, in the one-loop approximation \eqref{dilatation} to the dilatation operator. We found it convenient to discuss a further parameter, the angle $\phi$ breaking the periodic boundary conditions \eqref{twisted} while preserving integrability. In fact, this angle naturally appears in a certain deformation of the original AdS/CFT set-up, see the article \cite{chapDeform} by Konstantinos Zoubos. 

Let us now discuss what needs to be done to apply the Bethe ansatz to AdS/CFT. The first step is to extend the set of allowed operators from the $\alg{su}(2)$ sector of $X$ and $Z$ fields \eqref{singlespin} to the full, infinite set of ``spins''. The full one-loop magnet (see the article \cite{chapChain} by Joe Minahan) with $\alg{psu}(2,2|4)$ symmetry is also integrable, the Hamiltonian is known, and the Bethe equations may be derived. The main new feature is, due to the extra components, the need for the so-called {\it nested} Bethe ansatz technique. E.g.~in a magnet with $\alg{su}(n)$ symmetry, one needs $n-1$ Bethe ansatz {\it levels}. It again exists in both versions, coordinate and algebraic. The basic idea is beautiful, but the details are rather gruesome, and will not be discussed here. If you want to learn it, the best article I know is a lecture course by Sutherland \cite{Sutherland:1985aa}. It is worth going through for other reasons as well, as the article's main topic is the Hubbard model, which is closely related to the AdS/CFT system. You can also try his original article \cite{Sutherland:1975vr}, or again his book \cite{Sutherland:2004aa}, and in particular Appendix B.8 therein.

The second, much more difficult step is to extend the integrable spin chain beyond the one-loop level, and to connect the resulting equations to the string sigma model, cf.~the articles of section II of the review. Adding radiative corrections to the dilatation operator, see \eqref{dilatation} leads to {\it long-range} spin chains, see the article \cite{chapLR} by Adam Rej, which do not fit well into the standard framework of the quantum inverse scattering method. No equally nice general theory along the lines of the nearest neighbor spin chains exists. It should be stressed that taking into account a {\it finite} number of corrections to \eqref{dilatation} does not lead to an integrable model. Once we go beyond one-loop, we have to deal with the all-orders system, including the notoriously difficult wrapping interactions (see \cite{chapHigher}, and \cite{chapLuescher}). The biggest impediment to applying the two techniques we discussed to the full model is that we {\it neither} know the exact dilatation operator $\mathbf{D}$ in order to apply the coordinate Bethe ansatz, {\it nor} the generating Lax operator of the model in order to apply the algebraic Bethe ansatz. It somewhat contradicts the idea of making an ansatz for $|\psi \rangle$, if we do not know which operator equation (see \eqref{schroedinger} or \eqref{transferev}) is to be diagonalized.

That does not mean, however, that progress is impossible, as this review collection proves. We only briefly hinted at the beautiful picture behind the coordinate Bethe ansatz, where the down-spins are considered nearly free particles in the background of the up-spin vacuum. When one takes any one particle (= ``magnon'') with momentum $p_k$ around the chain, the standard phase factor $e^{i\,p_k\,L}$ of a would-be free particle gets modified by strictly pairwise (= ``factorized scattering'') collisions with all other particles carrying momentum $p_j$, modifying the phase with a two-body {\it S-matrix} element $S(p_k,p_j)$: Please take another look at \eqref{cbaeqs}. This idea works beautifully for AdS/CFT, and leads to the so-called {\it asymptotic Bethe ansatz}, see \cite{chapSMat} and \cite{chapSProp}, as well as the detailed discussion, using both coordinate and algebraic Bethe ansatz formalism, in the Ph.D.~thesis of Marius de Leeuw \cite{deLeeuw:2010nd}. The way it works is that one just assumes the existence of an integrable all-loop Hamiltonian, without actually knowing it exactly, and then fixes all S-matrix elements by symmetry, as well as further considerations such as crossing invariance.
Note that ``ansatz'' is now used in a slightly different fashion. The new principle is to ``make a guess for the solution, check self-consistency, and hope that it works''. And it actually only works up to the already mentioned wrapping interactions. The way around it are the somewhat empirical and arcane techniques discussed in \cite{chapLuescher} by Romuald Janik, \cite{chapTrans} by Volodya Kazakov and Kolja Gromov, and \cite{chapTBA} by Zoltan Bajnok of this collection. However, according to the thinking of this author, the following question remains open: Is there a non-asymptotic, {\it exact} Bethe ansatz for the  AdS/CFT system?

%%%%%%%%%%%%%%%%%%%%%%%%%%%%%%%%%%%%%%%%%%%%%%%%%%%%%%%%%%%%%%%%%%%%%%%%%%%%%%%%
\section{Bethe Equations without Ansatz: Q-Operator}
\label{sec:Baxter}

Somewhat ironically, this author wrote this review on the Bethe ansatz technique even though he no longer believes it to be the most elegant and powerful technique to solve a given quantum integrable model. Certain deformations of the XXX chain exist, where the matrix elements of the Lax operator are replaced by functions periodic (trigonometric XXZ, or 6-vertex model) model or double-periodic (elliptic XYZ, or 8-vertex model) in the spectral parameter.\footnote{%
To date these do not seem to play a major role in some AdS/CFT setting.}
The XYZ model was first solved in the early seventies by Baxter, introducing what is now known as the {\it Q-operator}. The limit of his construction back to the XXX case is very subtle. In fact, the Q-operator for the XXX chain with compact spin-$\frac{1}{2}$ representation (our illustrative example of this review) was only explicitly constructed very recently in \cite{Bazhanov:2010ts}. The main difference to Bethe's approach is that {\it no ansatz} is required to solve the model. To conclude this review, let us. therefore, just hint at the elegant and powerful way to derive the Bethe equations \eqref{betheeqs} using this method. Here the starting point is neither the Hamiltonian \eqref{hamiltonian} as for the coordinate Bethe ansatz nor the Lax operator \eqref{laxfaddeev} for the algebraic Bethe ansatz, but the ``generating objects'' are two novel Lax-operators 
\begin{equation}\label{Lplusminusu}
{\rm L}^-_l(u)=
\left( \begin{array}{cc}
1 \ \  & ~\mathbf{a}_-^\dagger\, \, \\
i\,\mathbf{a}_- \, \,  \,&u+i\,\mathbf{a}_-^\dagger \mathbf{a}_-
\end{array} \right)_l\, ,
\quad {\rm and} \quad
{\rm L}^+_l(u)=
\left( \begin{array}{cc}
u-i\,\mathbf{a}_+^\dagger \mathbf{a}_+ \, \, & ~~i\,\mathbf{a}_+^\dagger\, \, \\
- \mathbf{a}_+ \, \,&~~1\, \, \end{array} \right)_l
\ ,
\end{equation}
which also satisfy various Yang-Baxter equations. Here the $2 \times 2$ matrices act on some spin chain site $l$, the operators $\mathbf{a}_\pm$, $\mathbf{a}_\pm^\dagger$ are however not $\alg{su}(2)$ Lie-algebra generators, but {\it harmonic oscillator} operators: $[\mathbf{a}_\pm,\mathbf{a}_\pm^\dagger]=1$. Then one may prove that, in a sense made precise in \cite{Bazhanov:2010ts}, Faddeev's Lax operator in \eqref{laxfaddeev} factors into $\mathcal{L}_{a,l}(u) \sim {\rm L}^-_l(u)\,{\rm L}^+_l(u)$, where now the auxiliary space $a$ is given by the tensor product ${\cal F}_+ \times {\cal F}_-$ of the two copies of harmonic oscillators. The Baxter operators are then constructed in analogy with the transfer matrix operator of \eqref{transfertrace},\eqref{monodromy} as the trace in these Fock spaces of a monodromy matrix built from \eqref{Lplusminusu} (with $h_\pm=\mathbf{a}_\pm^\dagger \mathbf{a}_\pm$)
\begin{equation}\label{Baxteru}
\mathbf{Q}_\pm(u) \equiv \frac{e^{\pm \frac{\phi}{2} u}}{\mbox{Tr}_{{\mathcal F}_\pm}(e^{- i\, \phi\, h_\pm})}
\mbox{Tr}_{{\mathcal F}_\pm} \left(e^{-i \,\phi\, h_\pm }\, {\rm L}^\pm_L (u) \otimes \cdots \otimes {\rm L}^\pm_1 (u)\right)\, .
\end{equation}
They are operators on the spin chain space \eqref{quantumspace}. To illustrate it let us return once more to the discussion of the $L=2$ chain of Section \ref{sec:Intro}. It is easy to use \eqref{Baxteru} with \eqref{Lplusminusu} to compute, say, $\mathbf{Q}_-(u;\phi)$ in the spin chain basis \eqref{L2basis}:
\begin{equation}
e^{-\frac{\phi}{2}u}\left(
\begin{array}{cccc}
 1 & 0 & 0 & 0 \\
 0 &\, \, u+\frac{1}{2} \cot \frac{\phi }{2}\,\, &\,\, \frac{1}{2} \cot \frac{\phi }{2}+\frac{i}{2} & 0 \\
 0 & \frac{1}{2} \cot \frac{\phi }{2}-\frac{i}{2} \,\,& \,\,u+\frac{1}{2} \cot \frac{\phi }{2} & 0 \\
 0 & 0 & 0 & u^2+u\cot \frac{\phi }{2} +\frac{1}{2\sin^2 \frac{\phi}{2}}-\frac{1}{4}
\end{array}
\right)
\end{equation}
Diagonalizing this operator by going to the basis \eqref{L2beigenbasisalttwist}, we find
\begin{equation}
e^{-\frac{\phi}{2}u}\left(
\begin{array}{cccc}
 1 & 0 & 0 & 0 \\
 0 & u+\frac{1}{2} \cot \frac{\phi }{4} & 0 & 0 \\
 0 & 0 & u^2+u\cot \frac{\phi }{2} +\frac{1}{2\sin^2 \frac{\phi}{2}}-\frac{1}{4} & 0 \\
 0 & 0 & 0 & u-\frac{1}{2} \tan \frac{\phi }{4}
\end{array}
\right)
\end{equation}
We see that the eigenvalues $Q_-(u)$ of the $\mathbf{Q}_-$-operators take the form ``exponential times polynomial'', and the same is true for $Q_+(u)$:
\begin{equation}\label{Qpoly}
Q_-(u)
=
e^{-\frac{\phi}{2}u}\,\prod_{k=1}^M (u-u_k),
\qquad
Q_+(u)
=
e^{+\frac{\phi}{2}u}\,\prod_{k=1}^{L-M} (u-u_k).
\end{equation}
The roots of the polynomials are precisely the ones we found in the course of the discussion of the $L=2$ solutions of the twisted Bethe equations! See \eqref{betheeqs} and the discussion just below. And indeed, using the two $\mathbf{Q}$-operators it is easy to solve the XXX model in an entirely algebraic fashion, as one may derive the following {\it operator equations}:
\begin{eqnarray}
\label{func1}
2\,i\sin\frac{\phi}{2}\, u^L
&=&
\mathbf{Q}_+(u+\frac{i}{2})\,\mathbf{Q}_-(u-\frac{i}{2})-\mathbf{Q}_+(u-\frac{i}{2})\,\mathbf{Q}_-(u+\frac{i}{2}),\\
\label{func2}
2\,i\sin\frac{\phi}{2}\, \mathbf{T}(u)
&=&
\mathbf{Q}_+(u+i)\, \mathbf{Q}_-(u-i)-\mathbf{Q}_+(u-i)\,\mathbf{Q}_-(u+i).
\end{eqnarray}
It is straightforward to prove (no ansatz here!) that the eigenvalues of the $\mathbf{Q}_\pm$-operators must always be of the form \eqref{Qpoly}. Then it is an easy exercise to derive the Bethe equations \eqref{betheeqs} from \eqref{func1}. Furthermore, the transfer matrix and, therefore, through \eqref{HfromT} the Hamiltonian of the Heisenberg chain, follow from \eqref{func2}. Nice, no?

This entirely algebraic methodology generalizes to spin chains with $\alg{su}(n)$ \cite{Bazhanov:2010jq} as well as  $\alg{su}(n|m)$ symmetry \cite{Frassek:2010ga}, thereby bypassing the rather tedious nested Bethe ansatz technique. It will be interesting to see how to describe the eigenstates in this language. In any case I hope the method will also lift to the full AdS/CFT system.

%%%%%%%%%%%%%%%%%%%%%%%%%%%%%%%%%%%%%%%%
\phantomsection
\addcontentsline{toc}{section}{\refname}
\bibliography{chapters,intads}
\bibliographystyle{utphys}

\end{document}